\documentclass{elsart}

\usepackage{graphicx}

\usepackage{amssymb}

\begin{document}

\begin{frontmatter}

\title{Nuclear mass systematics using neural networks}

\author{S.~Athanassopoulos},
\ead{sathanas@cc.uoa.gr}
\author{E.~Mavrommatis\corauthref{cor}}
\corauth[cor]{Corresponding author.}
\ead{emavrom@cc.uoa.gr}
\address{Physics Department, Division of Nuclear \& Particle Physics, University of Athens, GR--15771 Athens, Greece}
\author{K.~A.~Gernoth}
\ead{K.A.Gernoth@umist.ac.uk}
\address{Department of Physics, UMIST, P.~O.~Box 88, Manchester M60 1QD, United Kingdom}
\author{J.~W.~Clark}
\ead{jwc@wuphys.wustl.edu}
\address{McDonnell Center for the Space Sciences and Department of Physics, Washington University, St.\ Louis, Missouri 63130, USA}

\begin{abstract}
New global statistical models of nuclidic (atomic) masses based on
multilayered feedforward networks are developed. One goal of such studies 
is to determine how well the existing data, and only the data, determines
the mapping from the proton and neutron numbers to the mass of the 
nuclear ground state. Another is to provide reliable predictive models 
that can be used to forecast mass values away from the valley of 
stability. Our study focuses mainly on the former goal and achieves 
substantial improvement over previous neural-network models of the 
mass table by using improved schemes for coding and training. The 
results suggest that with further development this approach may 
provide a valuable complement to conventional global models.
\end{abstract}

\end{frontmatter}

\def\sigmarms{$\sigma_{\mathrm rms}$}

\section{Introduction}
The problem of devising global models of nuclidic (atomic) masses 
has a long history, going back to the early work of Bohr, von
Weizs\"acker and Bethe based on the liquid drop model (see
refs.~\cite{1,2a,2b,2c,2d,Lunney} for reviews). The principal objectives 
are (i) a fundamental understanding of the physics of the mass 
surface and (ii) the prediction of the masses of ``new'' nuclides 
far from stability, both in the superheavy region and in the 
regions approaching the proton and neutron drip lines.

The actual predictions for the masses are of great current interest in
connection with present and future experimental studies of nuclei far
from stability, conducted at heavy-ion and radioactive ion-beam 
facilities \cite{3}. The results are also useful for such astrophysical
problems as nucleosynthesis and supernova explosions. The spectrum
of models of the atomic mass table ranges from those with high theoretical 
input that take explicit account of known physical principles in terms
of a relatively small number of fitting parameters, to models that 
are shaped mostly by the data and very little by theory and thus have 
a correspondingly larger number of adjustable parameters. 
Epitomizing models of the former class are the 
macroscopic/microscopic models of M\"oller, Nix, and coworkers
\cite{1,4,5,FRDM}, and the semi-microscopic models of Pearson, Tondeur, 
and coworkers \cite{Pearson1,Pearson2,Pearson3}.  The models of M\"oller 
et al. appeal to the macroscopic descriptions provided by the liquid 
drop and droplet models, solve a one-body Schr\"odinger 
equation to incorporate single-particle degrees of freedom, and 
include pairing through semi-microscopic calculation. A prominent
version, which sets the standard for state-of-the-art theory-based 
models, is the finite-range droplet model (FRDM) detailed in ref.~\cite{FRDM}.
The models of Pearson, Tondeur, and coworkers are based on the Hartree-Fock
method, with pairing correlations described by either a BCS or Bogolyubov 
treatment.  The current version, namely the HFB2 model \cite{Pearson3}, 
features the Bogoliubov approach and an improved Skyrme force. 
 
In this work we use neural networks to develop global nuclear mass
models which are situated far toward the other end of the spectrum, 
where one (in the ideal) seeks to determine the degree to which 
the entire mass table is determined by the existing experimental 
data, and only the data.  During the last decade, artificial neural 
networks have been utilized to construct predictive statistical 
models in a variety of scientific problems ranging from astronomy 
to experimental high-energy physics to protein structure 
\cite{Cherkassky,Clark_Ristig}.  In a typical application, 
a multilayer feedforward neural network is trained with 
back-propagation or some other supervised training algorithms 
\cite{Hertz,Clarkpmed,Haykin,Bishop} so as to create a ``predictive'' 
statistical model of a certain input-output mapping, which may 
in general be physical or mathematical in character.  Information 
contained in a set of learning examples of the input-output 
association is embedded in the weights of the connections between 
the layered units.  This information may (or may not) be sufficient 
to allow the trained network to make reliable predictions for 
examples outside the learning set.  At any rate, the network is taught
to generalize (well or poorly), based on what it has learned
from the set of examples.  In the more mundane language of function
approximation, the neural-network model provides a means for 
interpolation or extrapolation.

Nuclear physics offers especially rich territory for ``data mining''
with neural nets.  On the one hand, a huge collection of 
high-quality experimental data is available for diverse properties of
more than 2000 nuclides.  On the other, quantitative calculation of
some properties of some classes of nuclei presents difficult challenges
even for the best {\it ab initio} quantum-mechanical theories and
phenomenological macroscopic/microscopic models. To date, global 
neural-network models have been developed for the stability/instability 
dichotomy, for the atomic-mass table, for neutron separation energies, 
for spins and parities, for decay branching probabilities of 
nuclear ground states, and for $\beta$ decay half-lives 
\cite{Gazula,Gernoth_Prater,Neural_Networks_8,cpc88,Clark_Lindenau,Gernoth_Lindenau,MSc_thesis,Condensed_13,Condensed_15,Fission,Kalman}.

In the work to be described here, neural nets have been trained to predict 
the nuclear mass excess or ``defect'' $\Delta M$, continuing the program 
established in refs.~\cite{Gazula,Gernoth_Prater,cpc88,Condensed_15,Fission}. 
In Sec.~2, we outline the training methodology that has been applied and 
specify the data sets used in the modeling process.  The results are 
presented and discussed in Sec.~3.  Finally, Sec.~4 states the general 
conclusions of the current study and views the prospects for further 
successes and further improvements in statistical prediction of 
atomic masses.

\section{Design and training of neural-network models}
Our immediate tasks are to specify (i) the structure and unit-dynamics
of the networks that will be developed to model the mass data and (ii)
the algorithm for their training. We must
also specify (iii) the data sets to be utilized together with (iv) 
the schemes for encoding and decoding input and output data.

\subsection{Architecture and Dynamics}
A multilayer feedforward architecture is adopted, with various numbers
of hidden layers and distributions of units among layers. The
gross architecture of a given net is summarized in the notation
($I$--$H_1$--$H_2$--...--$H_L$--$O$)[$P$], where $P$ is the total
number of weight/bias parameters and $I$, $H_i$, and $O$ are integers
that indicate, respectively, the numbers of neuron-like units in the
input layer, the $i$th intermediate (or ``hidden'') layer, and the
output layer.  Unless otherwise indicated, each unit in a layer 
is connected to all units to the next layer.  The connection from 
unit $m$ to unit $n$ is characterized by a real-number weight 
$w_{mn}$ with initial value positioned at random in the range 
$[-1,1]$. 

When a pattern $\mu $ is impressed on the input interface, the activities of the input units are set in accordance to the coding scheme assumed (see section 2.4.). Each unit in a hidden layer or in the output layer receives a stimulus 
$u_n=\sum_m w_{mn} a_m$, where the $a_m$ are the activities of the units in the immediately
preceding layer.  The activity of generic unit $m$ in the hidden or output layers is in general a
nonlinear function of its stimulus, $a_m=g(u_m)$. In our work,
the activation function $g(u)$ is taken to have the logistic form,
$g(u) = \bigl[1 + \exp (-u) \bigr]^{-1}$. The system response may be decoded from the activities of the units of the output layer also in accordance to the coding scheme assumed. The dynamics is particularly simple: the states of all units within a given layer are updated
in parallel and the layers are updated successively, proceeding
from input to output.

\subsection{Training Algorithms}
Several training algorithms exist that seek to minimize the cost function with 
respect to the network weights. For the cost function we make the traditional choice 
of the sum of squared errors calculated over the learning set, or more specifically
\begin{equation}
E=\sum_\mu E^{(\mu )}=\frac 12\sum_{{\mu},i} 
(t_i^{(\mu)}-o_i^{(\mu)})^2 \label{sumofsquares} \,,
\end{equation}
where $t_i^{(\mu)}$ and $o_i^{(\mu)}$ denote, respectively, the
target and actual activities of unit $i$ of the output layer for 
input pattern (or example) $\mu$. In global modeling of the atomic 
mass table, the root-mean-square error \sigmarms~has been 
widely adopted as the key figure of merit, and we shall use
its value, calculated over various data sets, to assess the
performance of neural-network models. When evaluated over
the learning set, this quantity evidently coincides with 
$(2E/N_p)^{1/2}$, where $N_p$ is the number of training examples.  

The most familiar training algorithm is standard back-propagation 
\cite{Hertz,Haykin} (hereafter often denoted SB), according to 
which the weight update rule to be implemented upon presentation 
of pattern $\mu$ is
\begin{equation}
\Delta w_{mn}^{(\mu )}=-\eta \frac{\partial E^{(\mu )}}{\partial w_{mn}}
+ \alpha \Delta w_{mn}^{(\mu -1)} \,,
\label{standardupdate}
\end{equation}
where $\eta$ is the learning rate $(0<\eta <1)$, $\alpha$ is 
the momentum parameter $(0<\alpha <1)$, and $\mu -1$ is the 
pattern impressed on the input interface one training step earlier.
The second term on the right-hand side of Eq.~(\ref{standardupdate}), 
called the momentum term, serves to damp out the wild oscillations 
in weight space that might otherwise occur during the gradient-descent 
minimization process that underlies the back-propagation algorithm.  
In our implementation of the SB algorithm, the learning rate and 
momentum parameters remain constant at $0.5$ and $0.9$ respectively. 

Our artificial neural networks are trained with a modified version
of the SB algorithm \cite{MSc_thesis,Condensed_15,Fission} that we have 
found empirically to be advantageous in the mass-modeling problem. In this 
new algorithm, to be denoted MB, the weight update prescription 
corresponding to Eq.~(\ref{standardupdate}) reads 
\begin{equation}
\Delta w_{mn}^{(\mu )}=-\eta \frac{\partial E^{(\mu )} }{\partial w_{mn}}
+\alpha S_{mn}^{(\mu -1)} \,, 
\label{MBupdate} 
\end{equation}
the momentum term being modified through the quantity
\begin{equation}
S_{mn}^{(\mu-1 )}=\frac{S_{mn}^{(\mu -2)}e+\Delta w_{mn}^{(\mu-1)}}{e+1} 
\,.
\label{Sterm}
\end{equation}
In the latter expression, $e$ is the number of the current
epoch, with $e=0,1,2,3, \ldots $.
An epoch consists of $N_p$ pattern presentations, the patterns being
chosen at random from the set of $N_p$ training examples.  
Many epochs of training are required to achieve acceptable
performance in the problem at hand. 

The variable $S^{(\mu -1)}_{mn}$ entering Eq.~(\ref{MBupdate}) is 
initialized as follows: For the first pattern to be presented in 
epoch $e=0$, it is set equal to zero. At the beginning of each new 
epoch ($e>0$), it is taken equal to the value reached by $S_{mn}$ at the 
end of the immediately preceding epoch. The replacement of 
$\Delta w_{mn}^{(\mu -1)}$ by $S_{mn}^{(\mu -1)}$ in the update 
rule for the generic weight $w_{mn}$ allows earlier patterns of the 
current epoch to have more influence on the training than is the
case for standard back-propagation.  By the time $e$ becomes
large, $S_{mn}^{(\mu -1)}$ is effectively zero.  It can be shown,
after rather lengthy algebra, that if a plateau region of the 
cost surface has been reached (i.e. $\partial E/\partial w_{mn}$ 
remains almost constant) and $e$ is relatively large, then 
Eq.~(\ref{MBupdate}) converges to 
\begin{equation} 
\Delta w_{mn}=-2\frac{1}{1-\alpha}\frac{\partial E} {\partial w_{mn}}  
\,,
\label{MBconverge}
\end{equation}
thus achieving an effective learning rate twice that of 
the SB algorithm (cf.~\cite{Hertz}).

\begin{figure*}
\begin{center}
\includegraphics[scale=.8]{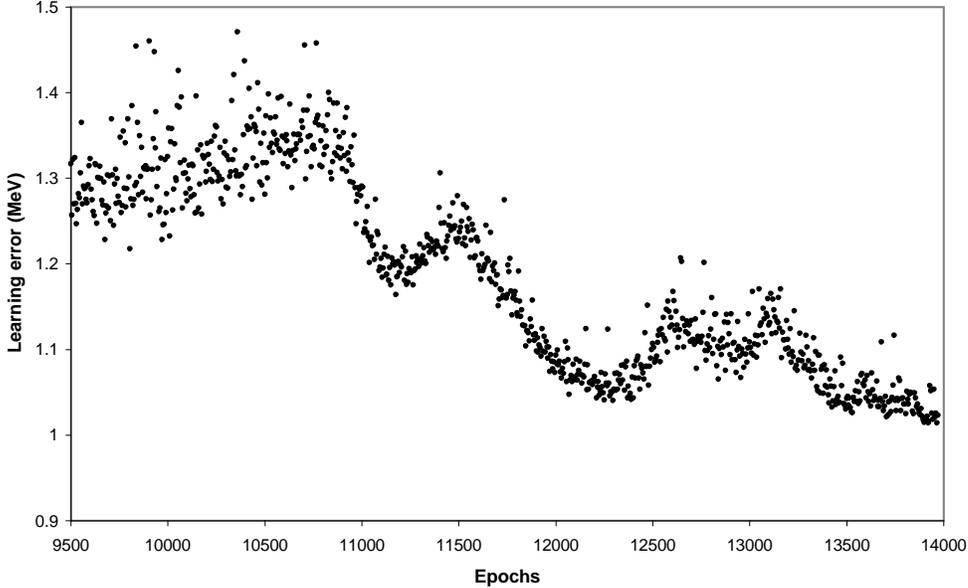}
\end{center}
\caption{\label{fig:figure1}Time course of the learning error (\sigmarms) for a network trained with
the modified backpropagation (MB) algorithm defined by eqs. (3) and (4) (typical example).}
\end{figure*}

In training neural networks, there is always the question of when
to stop the process. If the network is trained for too short a
period, the training data will not be adequately fitted. On the
other hand, if the training period is continued for too long,
generalization (i.e. prediction) will suffer, since the ``overtrained''
network will be specialized to the idiosyncrasies of the particular
learning set that has been supplied. Thus some reasonable compromise
must be struck between the {\it desiderata} of a good fit and 
good prediction.
In our computer experiments, we have adopted the following criterion.
A given training run consists of a relatively large number of epochs, 
specified beforehand. During such a run, we monitor not only the cost 
function for the patterns in the learning set, but also for a separate 
{\it validation set} of nuclei whose masses are known. The ``trained'' 
network model resulting from a given run is taken as that network with the 
set of connection weights producing the smallest value of the cost 
function on the validation set, over the full course of the run. While 
the members of the validation set are {\it not} used in the weight 
updates of the MB (or SB) training rule, they clearly do 
affect the choice of model. Therefore, accuracy on the validation set 
cannot strictly be regarded as a measure of predictive performance, 
although in practice it may still provide a useful indicator 
of this aspect of the model. To obtain a clean measure of predictive 
performance, still a third set of examples is needed:  a {\it test (prediction) set} 
that is never referred to during the training process.  

Another general problem that one faces in training neural networks is
that of the optimal architecture in terms of the numbers of layers 
and the numbers of units in each layer. As in most neural-network
applications, we have simply followed a ``trial-and-error'' approach 
to this problem; certainly no claim can be made that a full 
optimization has been achieved.

Numerical experiments have shown 
that performance can be improved by modifying the learning rate $\eta$ 
and momentum parameter $\alpha$ during training.  In applying the 
MB algorithm, $\eta$ and $\alpha$ are assigned 
starting values 0.5 and 0.9, respectively, and the validation 
error is calculated every five epochs. If this error decreases 
for two or more consecutive evaluations, the learning rate $\eta$
(with $0<\eta <3)$ is increased by 0.02; otherwise it is decreased 
by 0.005. The momentum parameter $\alpha$ is usually set to $0.9999$ 
when $e$ becomes relatively large.  Comparative studies of the 
mass-modeling problem (to be summarized in Table 1 of Section 
3) demonstrate that this training procedure, in conjunction
with the MB update rule (\ref{MBupdate})--(\ref{Sterm}), generally yields 
better results than does the SB algorithm.

The evolution of the learning error under application of the MB algorithm
is illustrated by the sample shown in fig.~1. We emphasize that this algorithm
departs from gradient descent, allowing the network to escape from local minima.

\subsection{Data Sets}
In exploring the prospects for statistical modeling of nuclear mass 
excesses, we have primarily employed a database O+N made up of
(i) $1323$ ``old'' (O) experimental masses which the 1981 M\"oller-Nix 
theoretical model \cite{4} was designed to reproduce, together 
with (ii) $351$ ``new'' (N) experimental masses, measured subsequently
for nuclei that lie mostly beyond the edges of the 1981 data collection 
when viewed in the $N-Z$ plane.  As discussed in ref.~\cite{1}, 
the O and N data sets were selected as part of a strategy
for quantifying the extrapolation capability (``extrapability'')
of global mass models -- i.e., their ability to predict the
atomic masses for nuclides far from stability.  These sets have
also been used in evaluating the predictive performance of 
neural network models of the atomic mass function, with set O
providing the fitting data (learning set) and set N the target
data for prediction (test set) (e.g., see ref.~\cite{Gernoth_Prater}).

To further characterize the interpolation/extrapolation capability 
of our models, we have also employed two data sets of $1303$ (M1) 
and $351$ (M2) nuclei and their masses, chosen {\it randomly} from the 
union of the O and N sets, after excluding 20 nuclides with poorly 
measured masses.  Together, these $1654$ cases form the database fitted
by the FRDM parametrization of ref.~\cite{FRDM}, i.e., by the best 
of the mid-90's theoretical models developed by the Los Alamos-Berkeley 
Group.  We also make use of another set of $158$ nuclei (denoted NB) that 
lie outside the O and N databases, the experimental masses of these 
examples being drawn from the NUBASE evaluation of nuclear and decay 
properties \cite{2b}. The locations of the nuclei of the O, N, and, 
NB sets in the $N-Z$ plane are shown in fig.~2. 

\begin{figure*}
\begin{center}
\includegraphics[scale=.8]{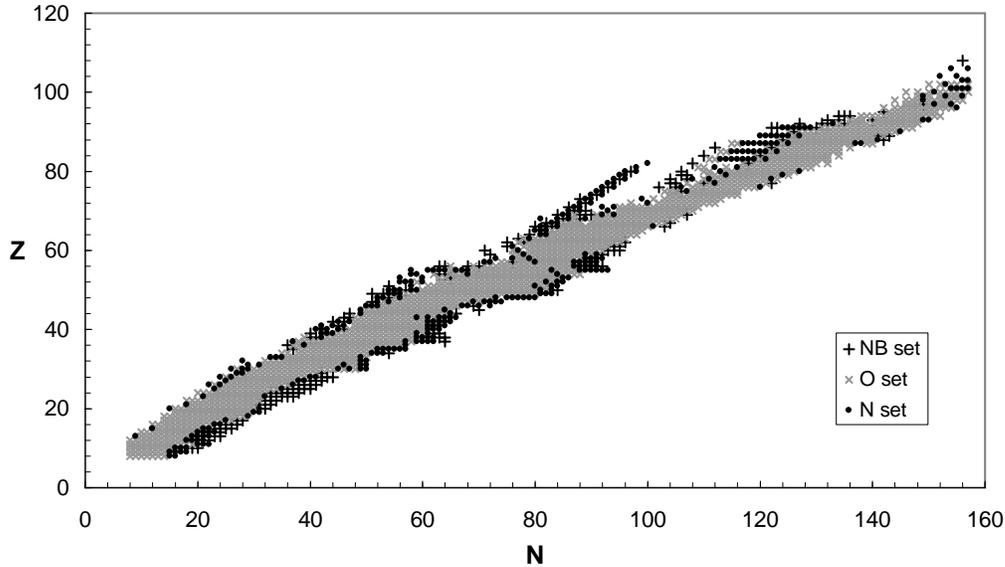}
\end{center}
\caption{\label{fig:figure2}Locations in the $N-Z$ plane are indicated for the 
O, N, and, NB data sets employed in neural-network modeling of nuclear 
mass excesses (see section 2.3).}
\end{figure*}

\subsection{Coding at Input and Output Interfaces}
We have considered several input coding schemes designed to 
facilitate learning of quantal properties (pairing, shell structure) 
that depend on the integral nature of $Z$ and $N$ (see refs. 
\cite{Gazula,Gernoth_Prater,MSc_thesis}).  The scheme that achieves
this aim most efficiently while keeping the number of weights
to a minimum is one that implements analog (floating-point) coding 
of $Z$ and $N$ in terms of the inputs of only two dedicated 
analog input neurons, which, however, are aided by two further 
binary (``on-off'') input units that encode the parity 
(even or odd) of $Z$ and $N$ \cite{MSc_thesis}.  The analog input 
units scale the $Z$ and $N$ values to the interval [0,1] in such 
a manner that the stimuli received by the logistic units in hidden 
and output layers remain within their best dynamical range. A less 
efficient scheme, introduced in ref.~\cite{Gazula}, utilizes banks 
of on-off units to represent $Z$ and $N$ as binary integers.

The mass excess computed by the network is represented by the activity of 
a single analog output unit. For the same reason as for the input units 
representing $Z$ and $N$, the target mass excess values $\Delta M$ 
are also scaled to the interval [0,1]. 

Several prescriptions have been tried for scaling the $Z$, $N$, 
and $\Delta M$ variables, two of which are represented in the
results reported here.   Extensive comparative studies of the MB and
SB training algorithms were based on the P1 prescription, which 
admits the ranges $[0,110]$, $[0,160]$, and $[-110,130]$ for the 
variables $Z$, $N$, and $\Delta M$, respectively.  In later work
we adopted the P2 prescription, which allows for the extended 
respective ranges $[0,130]$, $[0,200]$, and $[-110,250]$, thereby 
providing ample room for new nuclei far from the stable valley.
The latter scaling recipe usually gives better results.

\section{Results}
We first present results of a comparative study of the quality of models 
generated with the modified back-propagation training algorithm MB and with 
the standard back-propagation routine SB. Seven pairs of models were 
constructed, one member of each pair being trained with MB and the other 
with SB, and both members of the pair being started from the same choice 
among seven different sets of random initial weights. All of these models 
have architecture ($4$--$10$--$10$--$10$--$1$)[281] and employ the 
P1 scaling recipe. In all seven cases, the values of the error measures
\sigmarms~attained by the MB algorithm for the learning, 
validation, and test sets are consistently smaller than the 
corresponding values achieved with SB.  A similar pattern is 
expected to hold for the P2 scaling prescription.

\begin{table}
\caption{Performance comparison of standard back-propagation (SB) 
and modified back-propagation (MB) training algorithms in the task of 
mass modeling. Results for the rms error \sigmarms~on learning, 
validation, and test sets are given for seven pairs of models 
with network architecture ($4$--$10$--$10$--$10$--$1$)[281]. The
members of each model pair belong to the same choice among seven initial 
sets of random weights. All models have been trained for a total 
of 20,000 epochs using the P1 scaling recipe defined in Section 2.4.} 
\begin{center}
\begin{tabular}{ccccc} 
\hline\hline
Set & Algorithm  & 
\shortstack{learning   (O) set  \\ \sigmarms (MeV) } & 
\shortstack{validation (N) set  \\ \sigmarms (MeV) } &
\shortstack{test       (NB) set \\ \sigmarms (MeV) } \\
\hline
1   & MB       	&	0.78  &	1.92 	 & 2.84 				\\ 	
    & SB		&	1.27 	&     2.57 	 & 3.12 				\\ \hline	
2   & MB		&	0.78 	&	1.93 	 & 2.34 				\\ 	
    & SB	 	&	1.11 	& 	3.73 	 & 4.53 				\\ \hline	
3   & MB		&	0.71 	&	1.25 	 & 1.88 				\\  	
    & SB		&	0.92 	& 	1.58	 & 2.44 				\\ \hline	
4   & MB		&	0.98 	&	1.81 	 & 3.10 				\\ 
    & SB		&	1.31 	& 	3.88	 & 4.36 				\\ \hline	
5   & MB		&	0.69 	&	1.54 	 & 2.14	   			\\  	
    & SB		&	1.32	& 	2.06	 & 2.79				\\ \hline	
6   & MB		&	0.66 	&	1.65 	 & 2.73 				\\ 	
    & SB		&	0.96	& 	1.81 	 & 3.03  				\\ \hline
7   & MB		&	0.62	&	1.59	 & 2.01 				\\  	
    & SB		&	0.98 	& 	3.71	 & 3.97            	 	\\ 	
\hline\hline
\end{tabular}
\end{center}
\end{table}

Specializing to the MB algorithm, we have carried out a substantial 
number of computer experiments for networks with various architectures 
and for networks with the same architecture but different random 
choices of initial weights. 

Performance measures of some of our best models (marked with asterisks) 
are reported in Table 2, together with similar performance data for two 
neural-network models from earlier studies and for three high-quality 
theoretical models, two from M\"oller et al.~\cite{1,FRDM} and 
one due to Pearson et al.~\cite{Pearson3}. The table entries are 
separated into two groups by a pair of horizontal lines.

\begin{table}
\caption{Performance measures of global models of the atomic mass
table derived from neural-network technology and from conventional 
theory and phenomenology (see text for details). Data sets are 
indicated in parentheses. Those rms error measures referring
to actual predictions, rather than fits, are printed in italic font.}
\begin{center}
\begin{tabular}{cccc} 
\hline\hline
\shortstack{Network architecture\\($I$--$H_1$--$H_2$--...--$H_L$--$O$)[$P$]} & 
\shortstack{learning set   \\ \sigmarms (MeV) } &
\shortstack{validation set \\ \sigmarms (MeV) } &
\shortstack{test set       \\ \sigmarms (MeV) } \\
\hline
M\"oller et al. (ref.~\cite{1})&	0.67 (O)	&	--	&	{\it 0.74} (N)	 	\\ \hline 	
\shortstack{($18$--$10$--$10$--$10$--$1$)[421] (ref.~\cite{Gernoth_Prater})	 \\$Z$ \& $N$ in binary \\$A$ \& $Z-N$ in analog}	
						&	0.83 (O)	&	--	&{\it 5.98}	(N)	 		\\ \hline
($4$--$40$--$1$)[245] (ref.~\cite{Kalman}) 
						&	1.07	(O) &	--	&	{\it 3.04} (N) 			\\ \hline
\shortstack{($4$--$10$--$10$--$10$--$1$)$^{*}$[281]	\\ $Z$ \& $N$ in analog and parity}
						&	0.71 (O)	&	2.28 (NB)	&	{\it 2.16} (N) 	\\ \hline
Support Vector Machine (ref.~\cite{Li}) &	0.70	(O) 	&	--	&	{\it 0.75} (N) 		\\ \hline\hline
M\"oller et al. (ref.~\cite{FRDM})	&	0.68  (M1)	&	0.71 (M2)	&	{\it 0.70} (NB)	\\ \hline 
Pearson et al. (ref.~\cite{Pearson3})&    0.67  (M1)	&	0.68 (M2)	&  0.73  (NB)		\\ \hline  	 	
\shortstack{($4$--$10$--$10$--$10$--$1$)$^{**}$[281]	\\ $Z$ \& $N$ in analog and parity}
						&	0.41 (M1)	& 0.47	 (M2)	&	{\it 1.48} (NB) 	\\ \hline 
\shortstack{($4$--$10$--$10$--$10$--$1$)$^{***}$[361]	\\ $Z$ \& $N$ in analog and parity}
						&	0.44 (M1)	& 0.44	 (M2)	&	{\it 0.95} (NB) 	\\ 
\hline\hline
\end{tabular}
\end{center}
\end{table}

The generalization ability (extrapability) of models belonging to 
the first group, whether statistical or theoretical, was assessed by 
treating the N data set as a test set. In this way, results obtained
by the neural-network approach could be directly compared with 
the results \cite{1} of the extrapability study carried out by
the Los Alamos-Berkeley group for the FRDM approach (and especially 
with the rms error values shown in the first row of Table 2).  
The first of the network models selected from earlier studies is of 
the five-layer architecture ($18$--$10$--$10$--$10$--$1$)[421]; 
it was constructed by Gernoth et~al.~\cite{Gernoth_Prater} using standard back-propagation, with 
binary encoding of $Z$ and $N$ and ``redundant'' analog encoding of 
the atomic mass number $A$ and the neutron excess $N-Z$. The three-layer 
network ($4$--$40$--$1$)[245] is due to Kalman, who adopted analog 
coding of $Z$ and $N$ and auxiliary parity units for these variables. 
In training this model, the input patterns were pre-processed 
by singular-value decomposition and the cost function minimized by
a Powell-update conjugate-gradient algorithm (for additional details, 
see refs.~\cite{cpc88,Kalman}).

The network model labeled with one asterisk was one of those created
mainly for evaluating the modified training procedure MB.  It has 
the five-layer architecture ($4$--$10$--$10$--$10$--$1$)[281] and 
employs parity-aided input coding and the scaling recipe P1.  
Utilization of the NB data for validation gives this model an 
{\it a priori} advantage over the earlier network 
models, for which only the learning set was involved in the training 
process. This advantage is clearly realized in practice.

Also included in the first group is a set of results \cite{Li} obtained recently with a Support
Vector Machine \cite{vapnik,cristianini}. 

The network models appearing in the second group employ the scaling recipe P2 and were trained with the 
mixed data sets M1 and M2 described in section 2.3. Thus, the training 
and validation examples include, in this case, members from both the 
O and N data sets.  The intent was to develop statistical models 
that can be compared more directly with the most refined FRDM model 
of ref.~\cite{FRDM}, recalling that the parameters of this model 
were fit to the 1654 examples of the M1+M2 database. The network model marked with two asterisks 
has the same five-layer architecture and number of weight parameters as the (*) network.
In the network model marked with three asterisks, 
we chose to introduce connections from the analog input units to 
all units of all the hidden layers, to avoid or reduce the degradation 
of information as it propagates toward the output unit. However, 
this innovation comes at the expense of increasing the number 
of weight parameters from the $281$ of the (**) network to $361$. 
The resulting system is the best neural-network model of 
the mass table yet achieved, based on the accuracy of the estimated 
masses of the NB set of nuclides, taken as the test set. The 
corresponding rms error is 0.95 MeV, which is to be compared with 
the figure 0.70 MeV obtained in the FRDM evaluation.  Also included 
in the third group are the corresponding rms errors for the HFB2 model 
\cite{Pearson3}. The parameters of this model have been adjusted 
to an extended data set of $1888$ nuclei, which however 
includes the $158$ nuclei of the NB set.   

\begin{figure}
\begin{center}
\includegraphics[scale=.6]{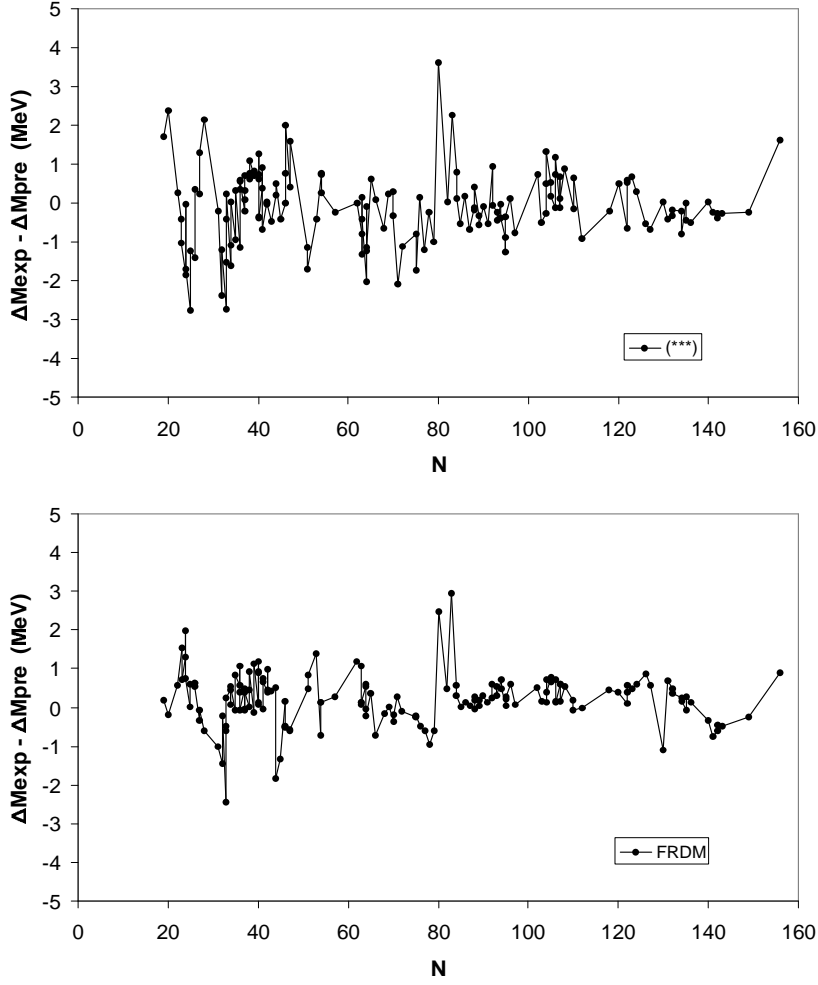}
\end{center}
\caption{\label{fig:figure3}Top panel: Deviations from experiment 
(in MeV) of mass-excesses values predicted by the neural-network model 
($4$--$10$--$10$--$10$--$1$)$^{***}$ for the NUBASE (NB) nuclei 
identified in fig.~2. The plot represents a projection of 
the mass surface onto a plane of constant $Z$ and thus shows dependence 
on neutron number $N$.  Bottom panel: Same for the FRDM 
evaluation \cite{FRDM}.}
\end{figure}

Further information on the performance of the (***) network is furnished 
in fig.~3. Here we compare the deviations from experimental data of the
mass-excess values generated by the net and by the FRDM evaluation, 
for the NB nuclei. The extrapolation capability of the (***) network  
model is better illustrated in fig.~4, which shows these deviations as a 
function of the number of neutrons away from the $\beta^{-}$ stability 
line.

\begin{figure}
\begin{center}
\includegraphics[scale=.6]{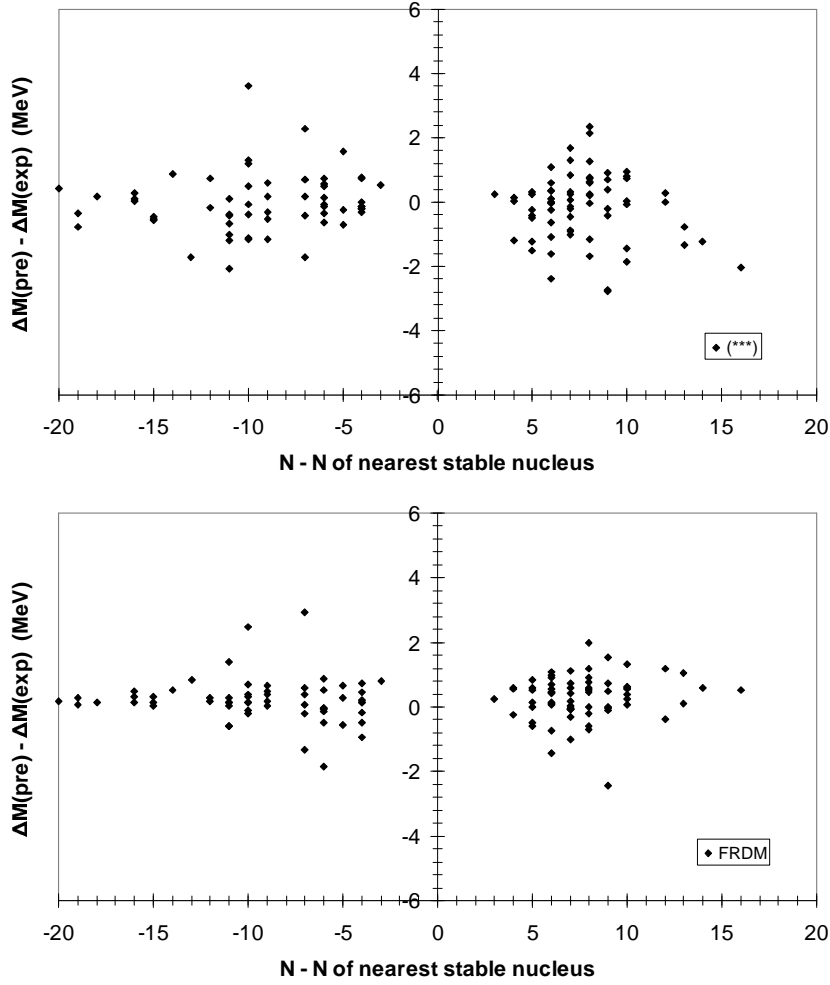}
\end{center}
\caption{\label{fig:figure4}Top panel: Deviations from experiment 
(in MeV) of mass-excesses values predicted by the neural-network model 
($4$--$10$--$10$--$10$--$1$)$^{***}$ for the NUBASE (NB) nuclei 
identified in fig.~2, as a function of the number of neutrons away 
from the line of $\beta^{-}$ stability. 
Bottom panel: Same for the FRDM evaluation \cite{FRDM}.}
\end{figure}

In spite of its residual shortcomings, the current generation 
of neural-network models of the mass table represents a significant 
step toward extrapability levels comparable with those reached 
by the best traditional global models rooted in quantum theory.
The ultimate test of any class of global mass models is the accuracy
that can be realized in the prediction of masses of nuclear species
prior to measurement. A text data file containing the mass-excess 
values predicted by the (***) network for 7709 nuclides 
($Z,N \geq 8$, $Z<120$, $N<200$) is available for downloading 
from {\em http://www.cc.uoa.gr/$\sim$sathanas/mass\_excess} 
(see file ``massfiles.txt'' for details). 
 
\section{Conclusions and prospects}
The present investigation is a continuation and an elaboration
of a research thrust 
\cite{Gazula,Gernoth_Prater,Neural_Networks_8,cpc88,Clark_Lindenau,Gernoth_Lindenau} 
that seeks to develop accurate global models of nuclear properties 
with demonstrable predictive power, within the arena of statistical 
methods based on multilayer neural networks. Our particular concern 
has been the central problem of modeling the systematics of nuclear
mass excesses. We have introduced a modified back-propagation 
algorithm (MB) along with new prescriptions for encoding and 
decoding input and output patterns.  The study has been based mainly 
on data sets selected by M\"oller and Nix \cite{1} for the purpose
of testing the extrapolation capabilities of global models of
atomic masses.  As seen in Table 2, our best network models 
of the nuclear mass excess $\Delta M$ display substantially improved 
performance relative to earlier attempts that use neural networks to 
predict masses far from the valley of $\beta$ stability.  A 
strong impetus for further improvement of this approach comes from
the production of new nuclei at radioactive beam facilities and 
heavy-ion colliders, as well as by the needs of supernova modeling 
and state-of-the-art theories of nucleosynthesis.

In closing, we would like to emphasize the conceptual and structural 
differences between 
\begin{itemize}
\item[(i)] statistical models of the atomic-mass function
constructed with the aid of learning rules operating purely on
the experimental data, without any overt imposition of physical
principles and theory, and
\item[(ii)]
the familiar theory-based phenomenological models, constrained
by the data. 
\end{itemize}
Although the latter models become more elaborate as the standards of 
description increase, they are nevertheless relatively compact,
having relatively few adjustable parameters, with transparent physical
meaning.  By contrast, the neural-network methodology is a more
abstract kind of ``engine'' that generates a statistical representation
of the experimental data having many parameters.  While this 
representation may have strong predictive power, its parameters 
are ordinarily (though not always) opaque to physical interpretation.  
In view of these fundamental differences, the two approaches 
should more fruitfully be viewed as complementary, rather than 
in competition.

We are currently exploring and implementing a number of refinements 
of neural-network approaches to the mass problem. These include 
the introduction of diverse pruning and network construction schemes and 
the application of other more powerful training (optimization) procedures.  
Additionally, we have made some initial attempts to construct an informative 
statistical model of the {\it differences} between the experimental 
mass-excess values $\Delta M^{\rm exp}$ and the theoretical values 
$\Delta M^{\rm th}$ given by the FRDM model of 
M\"oller et~al.~\cite{FRDM}.  This study is being pursued with 
the hope of revealing subtle regularities of nuclear structure not 
yet embodied in the best microscopic/phenomenological models of 
atomic-mass systematics.  To date, the results have not been
illuminating in terms of the emergence of systematic trends --
a tentative finding which, if sustained, could imply that the residual
physical corrections to the theoretical model are small but numerous, 
and of fluctuating size and sign.  Also under investigation is the 
potential of Support Vector Machines \cite{vapnik,cristianini} for 
systematic development of near-optimal statistical models of atomic 
masses and other nuclear properties.  The results reported in 
table 2 are suggestive of the power of this approach.

\section{Addendum}
After completing the training of the (***) network, the AME$03$ 
atomic mass evaluation \cite{Ame03} was published.  This compilation
made available precision mass measurements for nuclei farther off 
the stability line, while providing corrected mass-excess values 
for nuclei already used in our study.  The next generation of 
neural-network models will be trained using the AME$03$ data.  
Already, however, we can further appraise the extrapability performance 
of the (***) network, the best neural-network model of 
the mass table yet achieved, by making use of $529$ new nuclei 
included in the AME$03$ evaluation, which extend beyond the edges 
of the $1654$-nuclide set M1+M2 as viewed in the $N-Z$ plane.  A text 
data file containing the predicted mass-excess values for these 
$529$ additional nuclides is available for downloading from 
{\em http://www.cc.uoa.gr/$\sim$sathanas/ mass\_excess} (see 
file ``massfiles.txt'' for details). The resulting value of 
\sigmarms~for these nuclei is 1.03 MeV, which is to 
be compared with the figures 0.58 MeV and 0.67 MeV obtained in 
the FRDM and HFB2 evaluations.  When comparing these results,
it should be kept in mind that the parameters of the HBF2 model
have been adjusted by making use of an extended data set of 
$1888$ nuclei, which includes $255$ of the $529$ nuclides.   
\begin{ack}
This research has been supported in part by the U.~S. National Science 
Foundation under Grant Nos.~PHY-9900173 and PHY-0140316 and by the 
University of Athens under Grant No.~70/4/3309.
\end{ack}

\end{document}